\documentclass[letterpaper, 10 pt, conference]{ieeeconf}  

\usepackage[detect-all=true,separate-uncertainty=true, multi-part-units=single,range-phrase={-},product-units = single]{siunitx}
\usepackage{graphicx}
\usepackage{xcolor}
\usepackage{tikz} 
\usetikzlibrary{arrows,positioning,shapes.geometric}
\usetikzlibrary{intersections}
\usepackage{import}
\usepackage[caption=false,font=small]{subfig}
\definecolor{myblue}{RGB}{0,0,128}
\definecolor{myLightblue}{RGB}{213,229,255}
\definecolor{myDarkblue}{RGB}{55,113,200}
\usepackage{amsmath}
\usepackage{booktabs}
\usepackage{dcolumn}
\usepackage{adjustbox}
\usepackage{amssymb}
\usepackage{amsfonts}

\IEEEoverridecommandlockouts                              

\title{\LARGE \bf
A Modified da~Vinci Surgical Instrument for \\ \textcolor{black}{Optical Coherence Elastography with Deep Learning}
}

\author{
Maximilian~Neidhardt$^{1,2}$*, 
Robin~Mieling$^{1}$*, 
Sarah~Latus$^{1}$,  
Martin~Fischer$^{1}$,\\ 
Tobias~Maurer$^{3,4}$, 
and Alexander~Schlaefer$^{1,2}$
\thanks{$^{1}$Institute of Medical Technology and Intelligent Systems, Hamburg University of Technology, Hamburg, Germany {\tt\small maximilian.neidhardt@tuhh.de}}
\thanks{$^{2}$Interdisciplinary Competence Center for Interface Research at University Medical Center Hamburg-Eppendorf and Technical University of Hamburg, Germany}
\thanks{$^{3}$Martini-Klinik Prostate Cancer Center Hamburg-Eppendorf, Hamburg, Germany}
\thanks{$^{4}$Department of Urology, University Medical Center Hamburg-Eppendorf, Hamburg, Germany}
\thanks{The authors state no conflict of interest. This work was partially funded by the TUHH $i^{3}$ initiative and the Interdisciplinary Competence Center for Interface Research (ICCIR) supported by TUHH and UKE. 
}%
\thanks{* Both authors contributed equally.}%
}

\newif\ifcopyright
	\copyrighttrue

\begin{document}

\ifcopyright
{\LARGE IEEE Copyright Notice}
\newline
\fboxrule=0.4pt \fboxsep=3pt

\fbox{\begin{minipage}{1.8\linewidth}  

		Personal use of this material is permitted.  Permission from IEEE must be obtained for all other uses, in any current or future media, including reprinting/republishing this material for advertising or promotional purposes, creating new collective works, for resale or redistribution to servers or lists, or reuse of any copyrighted component of this work in other works. \\
		
		Published in: Proceedings of the 2024 10th IEEE RAS/EMBS International Conference for Biomedical Robotics and Biomechatronics (BioRob). IEEE, 2024.\\
  
        The final version of record is available at https://doi.org/10.1109/BioRob60516.2024.10719827 
		
\end{minipage}}
\else
\fi
\graphicspath{{./images/}}

\maketitle
\thispagestyle{empty}
\pagestyle{empty}

\begin{abstract}
Robot-assisted surgery has advantages compared to conventional laparoscopic procedures, e.g., precise movement of the surgical instruments, improved dexterity, and high-resolution visualization of the surgical field. However, mechanical tissue properties may provide additional information, e.g., on the location of lesions or vessels. While elastographic imaging has been proposed, it is not readily available as an online modality during robot-assisted surgery. We propose modifying a da~Vinci surgical instrument to realize optical coherence elastography (OCE) for quantitative elasticity estimation. The modified da~Vinci instrument is equipped with piezoelectric elements for shear wave excitation and we employ fast optical coherence tomography (OCT) imaging to track propagating wave fields, which are directly related to biomechanical tissue properties. All high-voltage components are mounted at the proximal end outside the patient. We demonstrate that external excitation at the instrument shaft can effectively stimulate shear waves, even when considering damping. Comparing conventional and deep learning-based signal processing results in mean absolute errors of \SI{19.27}{\kilo\Pa} and \SI{6.29}{\kilo\Pa} \textcolor{black}{ for a range of \SIrange{17}{139}{\kilo\Pa},} respectively. These results illustrate that precise quantitative elasticity estimates can be obtained. We also demonstrate quantitative elasticity estimation on ex-vivo tissue samples of heart, liver and stomach, and show that the measurements can be used to distinguish soft and stiff tissue types.
\end{abstract}
\section{INTRODUCTION}
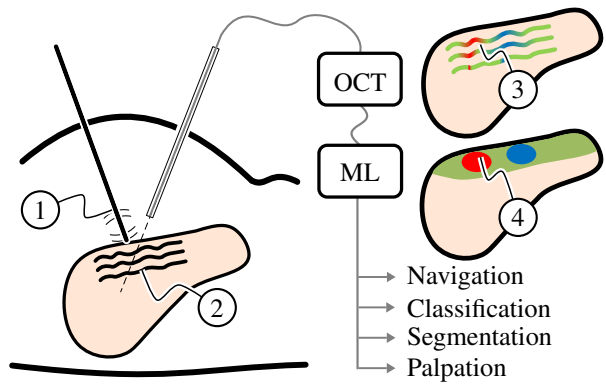
\begin{figure}[tb]
     \begin{tikzpicture}
         \node[anchor=south west,inner sep=0] at (0,0) {\includegraphics[width=0.95\linewidth]{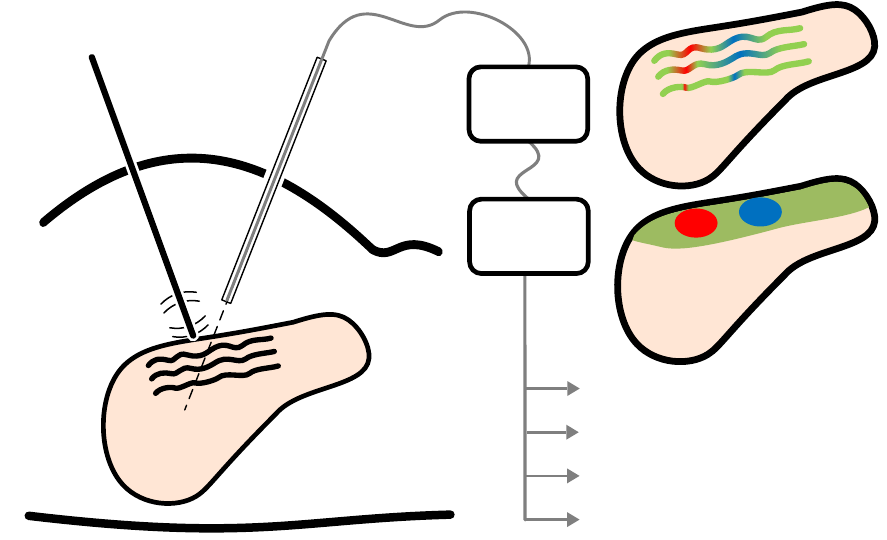}};
         
        \node at (4.9+0.5,1.4-0.41*0) [align=center, anchor=west]{Navigation};
        \node at (4.9+0.5,1.4-0.41*1) [align=center, anchor=west]{ Classification};
        \node at (4.9+0.5,1.4-0.41*2) [align=center, anchor=west]{ Segmentation};
        \node at (4.9+0.5,1.4-0.41*3) [align=center, anchor=west]{ Palpation};

        \draw [white, line width=0.8mm] plot [smooth] coordinates {(0.65,2.3) (1.2,2.5) (1.25,2.2) (1.6,2.3)};
        \draw [black, line width=0.2mm] plot [smooth] coordinates {(0.65,2.3) (1.2,2.5) (1.25,2.2) (1.6,2.3)};
        \node[shape=circle,draw,inner sep=2pt,thick, fill=white] at (0.65,2.3) (char) {1};
        
        \draw [white, line width=0.8mm] plot [smooth] coordinates {(3,1) (2.5,1.3) (2.2,1.1) (2,1.5)};
        \draw [black, line width=0.2mm] plot [smooth] coordinates {(3,1) (2.5,1.3) (2.2,1.1) (2,1.5)};
        \node[shape=circle,draw,inner sep=2pt,thick, fill=white] at (3,1) (char) {2};

        \draw [white, line width=0.8mm] plot [smooth] coordinates {(7.0,3.9) (6.8,4.4) (6.6,4.3) (6.5,4.5)};
        \draw [black, line width=0.2mm] plot [smooth] coordinates {(7.0,3.9) (6.8,4.4) (6.6,4.3) (6.5,4.5)};
        \node[shape=circle,draw,inner sep=2pt,thick, fill=white] at (7.0,3.9) (char) {3};
        
        \draw [white, line width=0.8mm] plot [smooth] coordinates {(7.0,2.2) (6.8,4.4-1.6) (6.6,4.3-1.6) (6.5,3)};
        \draw [black, line width=0.2mm] plot [smooth] coordinates {(7.0,2.2) (6.8,4.4-1.6) (6.6,4.3-1.6) (6.5,3)};
        \node[shape=circle,draw,inner sep=2pt,thick, fill=white] at (7.0,2.2) (char) {4};

        \node at (4.9,2.8) [align=center, anchor=center]{ML};
        \node at (4.9,4.0) [align=center, anchor=center]{OCT};

        
   \end{tikzpicture}
    
    \caption{
		\textbf{Minimal Invasive Elasticity Estimation:} 
        A mechanical stimulation (1) excites shear waves (2) which propagates inside the tissue. Tissue displacement due to the propagating waves is imaged with OCT. Wave field characteristics are dependent on the tissue properties as indicated in color (3). A machine learning network estimates quantitative elasticity values from image data to map soft tissue properties (4). The elasticity estimates could be further used  \textcolor{black}{for downstream tasks}.}
	\label{fig:Motivation}
\end{figure}

Robot-assisted minimally invasive surgery has advantages like improved precision, shortened recovery, and improved ergonomics for the surgeon~\cite{wee2020systematic}.
It has been widely adopted and the da~Vinci surgical system (Intuitive Surgical, Inc., USA) is the most frequently used platform for robot-assisted surgery (RAS)~\cite{Farinha.2022}. Typically, RAS is based on endoscopic vision, often using stereo cameras, which provide an excellent overview of the surgical field. However, the camera images only show the tissue surface and do not measure biomechanical tissue properties~\cite{Bandari.2020,Bergholz.2023}, e.g., its elasticity, which may provide additional information complementary to the visual appearance. Conventionally, palpation is used to identify tissues and lesions, e.g., in open surgery, and several elastographic imaging modalities have been developed, including ultrasound elastography and magnetic resonance elastography, which allow using tissue stiffness during diagnosis. However, intra-operative real-time quantitative elastography is currently not available, particularly during minimally invasive procedures. Yet, estimating tissue elasticity is beneficial for navigation, e.g., to identify the tissue type including lesions and vessels~\cite{kennedy2020diagnostic}, to detect the shape, e.g., of lesions, or even to generate force feedback \cite{Neidhardt.2023}. Note, that while surgeons can train to largely compensate for the lack of haptics and force feedback, more autonomous robot assistance may still benefit from elasticity estimates, e.g., as an additional safety feature. In this context, we consider a minimally invasive RAS scenario as illustrated in Fig.\ref{fig:Motivation}, where the instruments and the laparoscopic camera are controlled via robotic arms and quantitative elasticity values are predicted from optical measurements. 

\begin{figure*}[tb]
   \subfloat[CAD Design]{
        \input{Figures/cad}
        \label{fig:CAD1}
    }
   \hfill
    \subfloat[da~Vinci Robot]{
        \includegraphics[height=50mm]{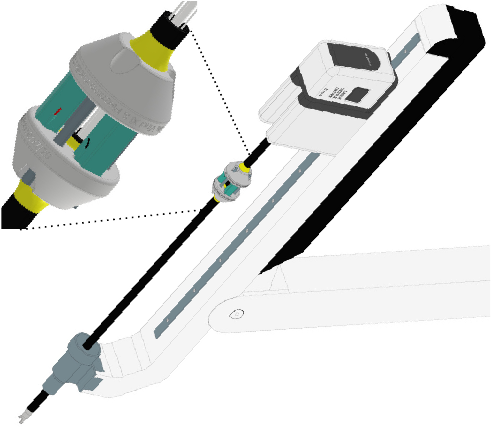}
        \label{fig:CAD2}
    }   
    \caption{\textbf{Modified da~Vinci Instrument:} (a) CAD design and image of the modified da~Vinci surgical instrument. The instrument is fitted with two 3D printed actuator mountings which hold three piezoelectric elements. The cables for instrument tip actuation are left unchanged and we preload the piezoelectric elements using three M3 screws. (b) The instrument can be inserted and operated with the da~Vinci robot as depicted.}
   \label{fig:CAD}
\end{figure*}

Optical coherence elastography (OCE) is particularly interesting for imaging subsurface structures in the context of RAS \textcolor{black}{ due to the high spatial and temporal resolution compared to magnetic or ultrasound based imaging} ~\cite{Kennedy.2014, Zvietcovich.2022}. Due to its non-invasive nature, optical elastography is safe and removes the need for direct contact between sensor and sample. In OCE, the external load is most commonly applied by excitation of shear waves on the surface of the soft tissue. By tracking the wave propagation in 2D, 3D, or even 4D images a precise quantification of tissue elasticity is possible~\cite{neidhardt20214d}. Note, that it is substantially more challenging to obtain quantitative estimates when solely tracking the robot position and the force between instrument and tip during palpation~\cite{mieling2023optical}. Tabletop OCE setups or OCE systems integrated into the clinical microscopes have already been demonstrated for ex-vivo or in-vivo tissue analysis, e.g. in ophthalmology~\cite{Zvietcovich.2022}. But for application in RAS, both, the image acquisition and the shear wave excitation must be performed locally on the tissue surface. While optical coherence tomography (OCT) imaging through a laparoscope has been proposed~\cite{Hariri.2009,Lee.2024}, shear wave excitation is complicated by the laparoscopic setting. In principle, shear waves can be excited using either piezoelectric elements, air-pulses, or acoustic radiation force~\cite{Zvietcovich.2022}. Piezoelectric elements are most commonly used because they offer a high load force, precise frequency tuning, and a wide bandwidth of excitation frequencies. \textcolor{black}{We propose placing the piezo elements outside the patient, at the base of the surgical tool. Thereby, we avoid high voltages inside the patient and modifications at the instrument tip which would hinder the operation of the tool. Hence, we investigate a novel approach using the da~Vinci instrument itself for shear wave excitation}.

We propose a modified da~Vinci instrument with integrated piezo actuators for shear wave excitation. Although the instrument is modified, we preserve all its surgical functions, i.e., the dexterous control of the instrument tip via the robot. All additional components are located at the proximal end of the instrument, outside the patient's body. In contrast to excitation concepts where the piezoelectric elements are placed at the tissue surface, we observe a more complex transmission of vibrations to the instrument tip. Consequently, the wave fields in the acquired OCT image data may be modulated. We therefore consider deep learning to directly correlate wave field patterns to the elasticity of soft tissue. In the following, we will first present our modified da~Vinci instrument for shear wave excitation. We will then validate its functionality on tissue-mimicking phantoms, ex-vivo tissue and inside a laparoscopic trainer. Finally, we demonstrate the benefits of deep learning for OCE in comparison to conventional elasticity estimation approaches.

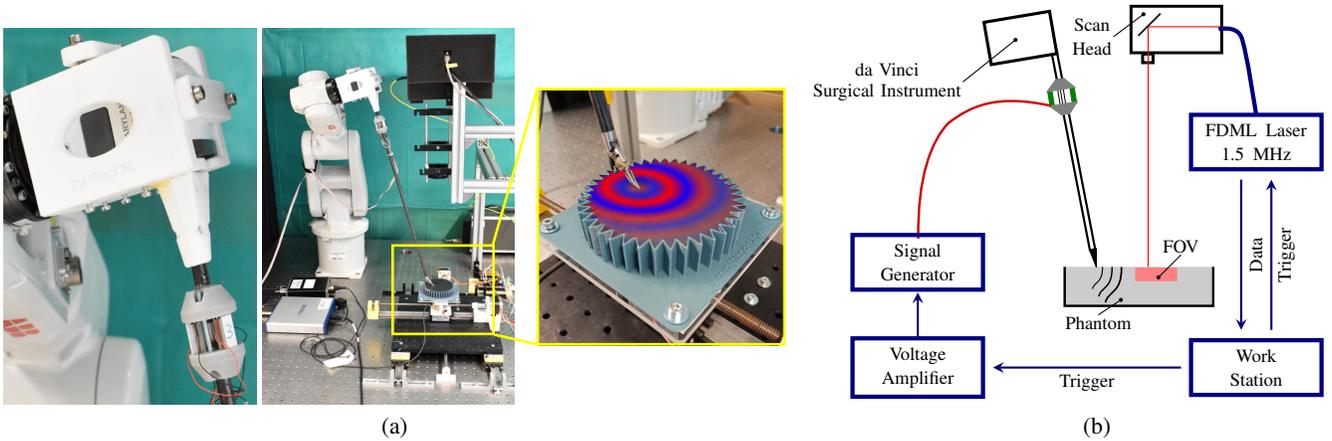
\begin{figure*}[tb]

    \subfloat[]{
        \includegraphics[height=50mm]{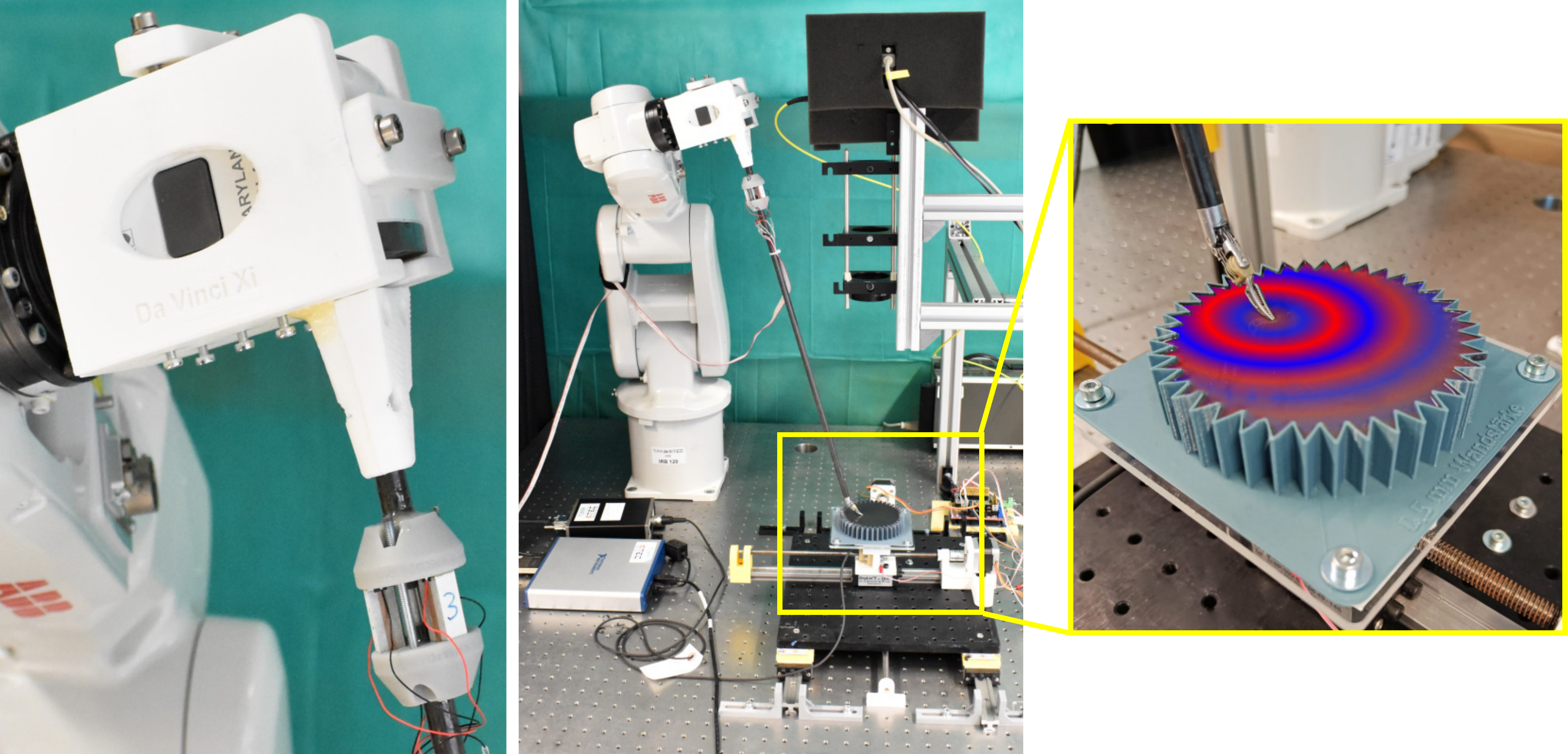}
        \label{fig:ExperimentalSetupA}
   }    
   \hfill
   \subfloat[]{
        \begin{tikzpicture}
        \node[anchor=south west,inner sep=0] at (0,0) {\includegraphics[height=55mm, trim={1.7cm 2cm 0 0},clip]{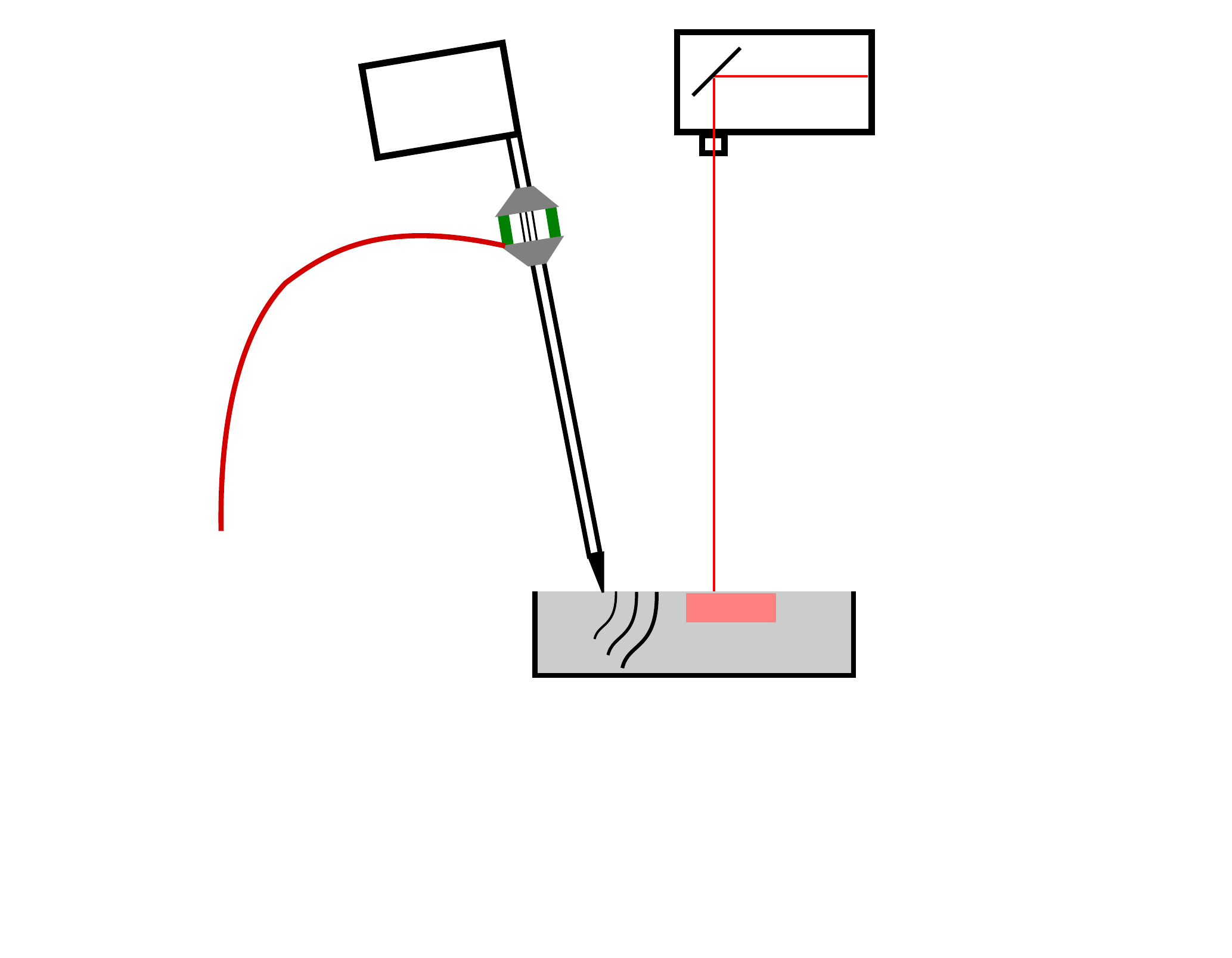}};
        \node[align=center,text width=15mm, draw=myblue, ultra thick] at (5+0.5,3.5) {\scriptsize FDML Laser \\[-1mm] 1.5~MHz}; 
        \node[align=center,text width=15mm, draw=myblue, ultra thick] at (1,1.9) {\scriptsize Signal \\[-1mm] Generator}; 
        \node[align=center,text width=15mm, draw=myblue, ultra thick] at (1,0.5) {\scriptsize Voltage \\[-1mm] Amplifier}; 
        \node[align=center,text width=15mm, draw=myblue, ultra thick] at (5+0.5,0.5) {\scriptsize Work \\[-1mm] Station}; 

        \draw [myblue, ultra thick] plot [smooth] coordinates {(5.5,3.85) (5.3,4.9) (5.0,5.0)};

        \draw[-stealth, thick, draw=myblue] (1,0.5+0.4) -- (1,1.9-0.4);
        \draw[-stealth, thick, draw=myblue] (5.2+0.5,0.5+0.5) -- node[rotate = 90,left=-2mm, align=center, anchor=center] {\scriptsize Trigger}   (5.2+0.5,3.5-0.5);
        \draw[-stealth, thick, draw=myblue] (4.8+0.5,3.5-0.5) -- node[rotate = 90,left=-2mm, align=center, anchor=center] {\scriptsize Data}   (4.8+0.5,0.5+0.5);
        \draw[-stealth, thick, draw=myblue] (5-0.5,0.5) --  node[above=-2mm, align=center, anchor=center] {\scriptsize Trigger} (2,0.5);

        \draw [fill=black] (3.8+0.2,4.5+0.7) circle (0.2mm); 
        \draw [] (3.8+0.2,4.5+0.7) -- (3.6,4.4+0.5)   node [align=center, left=-1mm] {\scriptsize Scan\\[-1mm] \scriptsize Head};
        
        \draw [fill=black] (3.5+0.2,1.3+0.1) circle (0.2mm); 
        \draw [] (3.5+0.2,1.3+0.1) -- (3+0.4,1+0.2)   node [below=-1mm] {\scriptsize Phantom};

        \draw [fill=black] (3.5+0.5+0.2,1.3+0.6-0.2) circle (0.2mm); 
        \draw [] (3.5+0.7,1.3+0.4) -- (3.5+0.3+0.7,1.3+0.5+0.2)   node [above=-1mm] {\scriptsize FOV};
        
        \draw [fill=black] (2.5-0.2,3.56+1+0.3) circle (0.2mm); 
        \draw [] (2.5-0.2,3.56+1+0.3) -- (1.6,4.3)   node [align=center, left=-1mm] {\scriptsize da~Vinci\\[-1.5mm] \scriptsize Surgical Instrument};
        
        
        \end{tikzpicture}
        \label{fig:ExperimentalSetupB}
       
   }
       \caption{\textbf{Experimental Data Acquisition:} (a) Experimental setup with a robot and the custom da~Vinci surgical instrument mounted to the robot's end-effector. Piezoelectric elements integrated inside the shaft of the da~Vinci instrument are excited to stimulate shear waves inside the tissue, as depicted in the image on the right. We image the propagating wave-field with high-speed OCT. (b) Sketch demonstrating data acquisition. A scanner attached to the OCT system acquires cross-sectional images with a temporal scan rate of \SI{11.4}{\kilo\hertz} with simultaneous wave excitation.} 
   \label{fig:ExperimentalSetup}
\end{figure*}

\section{MATERIAL AND METHODS}

\subsection{Surgical da~Vinci Instrument}
In order to realize shear wave excitation directly via the da~Vinci instrument, we integrate piezoelectric actuators into the instrument shaft. Thus, we can excite the instrument itself, and thereby the tip of the instrument. The design and the individual components of the modified da~Vinci instrument are depicted in Fig.\ref{fig:CAD1}. We integrate three piezoelectric elements (PSt 150/7x7/20, Piezomechanik GmbH, Germany) at the proximal end of the instrument. The proximal mounting ensures that all additional components are outside of the patient when operating the instrument with a robot, as depicted in Fig.\ref{fig:CAD2}. Each piezo has a ceramic cross-sectional area of \SI{7}{\milli\meter} \texttimes\ \SI{7}{\milli\meter}, a length of \SI{18}{\milli\meter}, a maximum stroke of \SI{28}{\micro\meter} and a maximum load force of \SI{4000}{\N}. We operate the elements with a voltage ranging from \SI{-25}{\V} to \SI{150}{\V}. Our prototype is manufactured in-house by (1) removing a \SI{10}{\milli\meter} section of the da Vinci tool's  \textcolor{black}{carbon fiber tube}, (2) two mounting plates are fixed to the \textcolor{black}{ carbon fiber tube} with epoxy adhesive, (3) the piezoelectric elements are placed between the mounting plates and tension is applied on the piezo elements with screws.  \textcolor{black}{During operation, the tool is mounted rigidly to the robot, the shaft is positioned inside a trocar and the grasper is in contact with the soft tissue.}

\subsection{Experimental Setup}
The experimental setup is depicted in Fig.\ref{fig:ExperimentalSetupA}. We use a high-speed swept-source OCT system (SS-OCT, OMES, Optores, Germany) to acquire one-dimensional depth-resolved scans with a temporal scan rate of \SI{1.5}{\mega\Hz}, a central wavelength of \SI{1315}{\nano\meter} and an axial resolution of \SI{15}{\micro\meter} in air. We use a scan head to acquire two-dimensional images with a temporal resolution of \SI{11.4}{\kilo\Hz} by deflecting the light beam with resonant oscillating mirrors. We measure the field of view (FOV) with a calibration phantom (R3L3S3P, Thorlabs, Germany) and report an effective spatial size of approximately \SI{3.5}{\milli\meter} \texttimes\ \SI{2}{\milli\meter} for an image size of 118 \texttimes\ 400~pixels along the lateral and depth axis. We design and print an adapter to mount the da~Vinci instrument to the robot's end-effector (IRB120, ABB, Switzerland) for streamlined data acquisition. The mounting includes screws that apply tension to the instrument's control gears to restrict the movement of the instrument tip during data acquisition.

\subsection{Data Acquisition and Data Pre-Processing}
We use gelatin phantoms as tissue surrogates and prepare batches of gelatin with a weight ratio of \SI{5}{\percent}, \SI{10}{\percent}, \SI{15}{\percent} and \SI{20}{\percent} gelatin to water. For each concentration, we manufacture 5 phantoms following the same recipe. During data acquisition, we place the samples on a planar motion stage as depicted in Fig.\ref{fig:ExperimentalSetupA}. The robot places the da~Vinci instrument on the gelatin sample. To ensure sufficient contact between the instrument tip and soft tissue we apply a maximum contact force of approximately \SI{100}{\milli\N} which is sensed with a force sensor mounted underneath the motion stage. Next, shear waves are excited iteratively at \SI{200}{\Hz}, \SI{400}{\Hz}, \SI{600}{\Hz}, \SI{800}{\Hz} and \SI{1000}{\Hz} while the instrument is kept stationary. At each frequency, a dataset is acquired containing 208 cross-sectional images, creating a 3D spatio-temporal representation of propagating shear waves. In total, we acquire datasets at 25 distinct positions on each gelatin phantom which results in a total of 500 acquired datasets per excitation frequency. Each position on the phantom is separated by \SI{5}{\milli\meter} to allow speckle augmentation. We perform uniform indentation tests on cylindrical gelatin samples to calculate the Young's modulus as our ground truth labels for training~\cite{neidhardt2022ultrasound}. We obtain  \SI{17}{\kilo\pascal}, \SI{56}{\kilo\pascal}, \SI{97}{\kilo\pascal} and \SI{139}{\kilo\pascal} for the four different gelatin concentrations, respectively. We pre-process each dataset by cropping 128~pixels beneath the sample surface. To reduce imaging noise, we apply a median filter with a kernel size of 3 \texttimes\ 3 \texttimes\ 3. 

\subsection{Laparoscopic Trainer and Soft Tissue}
\begin{figure}[tb]
    \centering
    
    \subfloat[]{
        
        \includegraphics[width=0.303\linewidth]{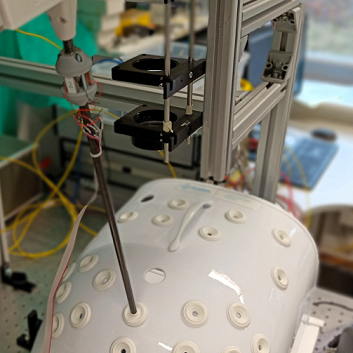}
        \label{fig:trainerA}
   }        
    \subfloat[]{
        \includegraphics[width=0.303\linewidth]{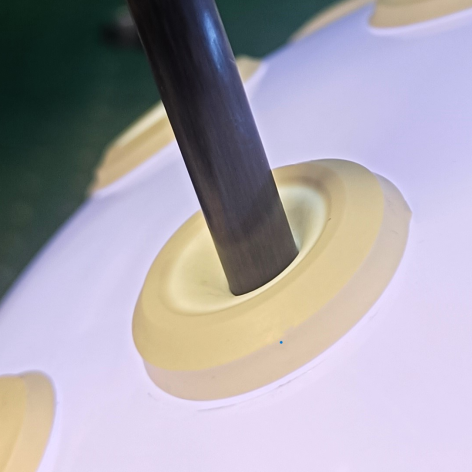}
        \label{fig:trainerB}
   }         
    \subfloat[]{
        \includegraphics[width=0.303\linewidth]{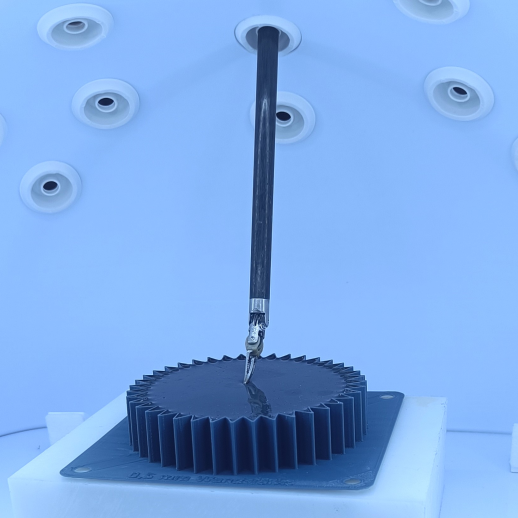}
        \label{fig:trainerC}
   }   
   \\
    \subfloat[]{
        \includegraphics[width=0.303\linewidth]{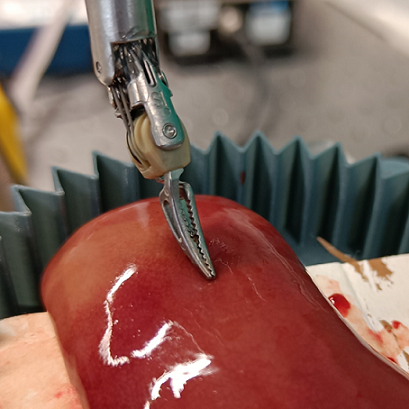}
        \label{fig:SoftTissueA}
   }   
   \subfloat[]{
        \includegraphics[width=0.303\linewidth]{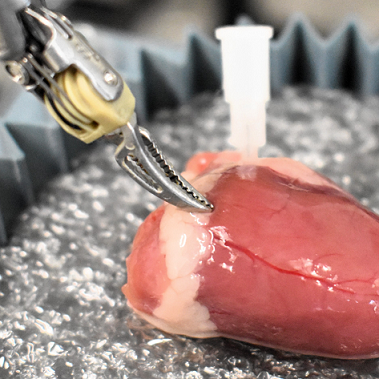}
        \label{fig:SoftTissueB}
   }    
   \hfill
  \subfloat[]{
        \includegraphics[width=0.303\linewidth]{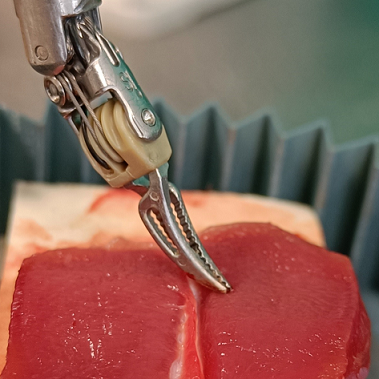}
        \label{fig:SoftTissueC}
   }
   
    \caption{
		\textbf{Laparoscopic Trainer and Soft Tissue Experiments:} We additionally validate wave excitation in a trainer that adds friction along the instrument shaft (a-c). Furthermore, we test our approach on ex-vivo (d) chicken liver, (e), chicken heart and (f) chicken stomach tissue. 
    }
    \label{fig:chicken_methods}
\end{figure}

Additionally, we record data from within a laparoscopic trainer (Kroton Box, Kroton Medical Technology, Poland) to estimate the effect of damping on the instrument's shaft. We record 20 datasets with and without damping of the laparoscopic trainer (Fig.\ref{fig:trainerA}-\ref{fig:trainerC}). In addition to the gelatin phantom data, we also evaluate our methods on ex-vivo chicken heart, liver, and stomach tissue (Fig.\ref{fig:SoftTissueA}-\ref{fig:SoftTissueC}). In total, we record 15 datasets at various positions for each soft tissue type.

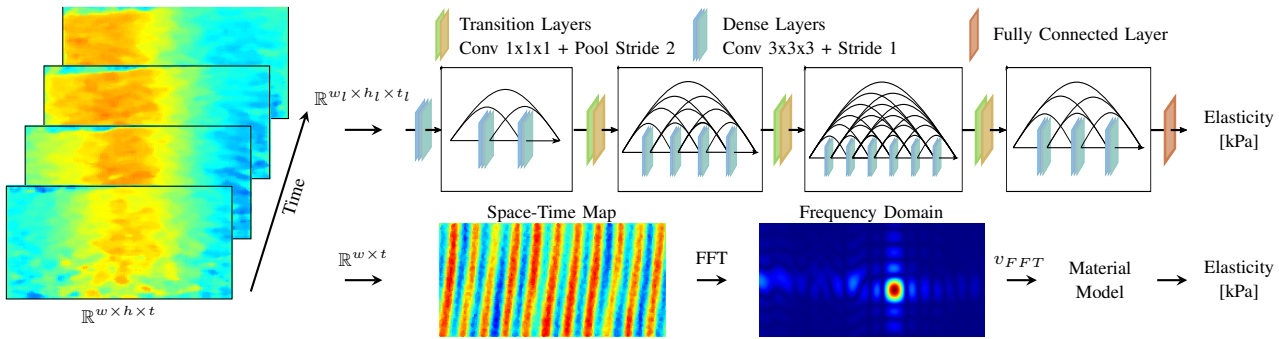
\begin{figure*}[tb]
        \begin{tikzpicture}
        \node[anchor=south west,inner sep=0] at (0.25*3,0.5+0.8*3) {\includegraphics[height=15mm,width=30mm]{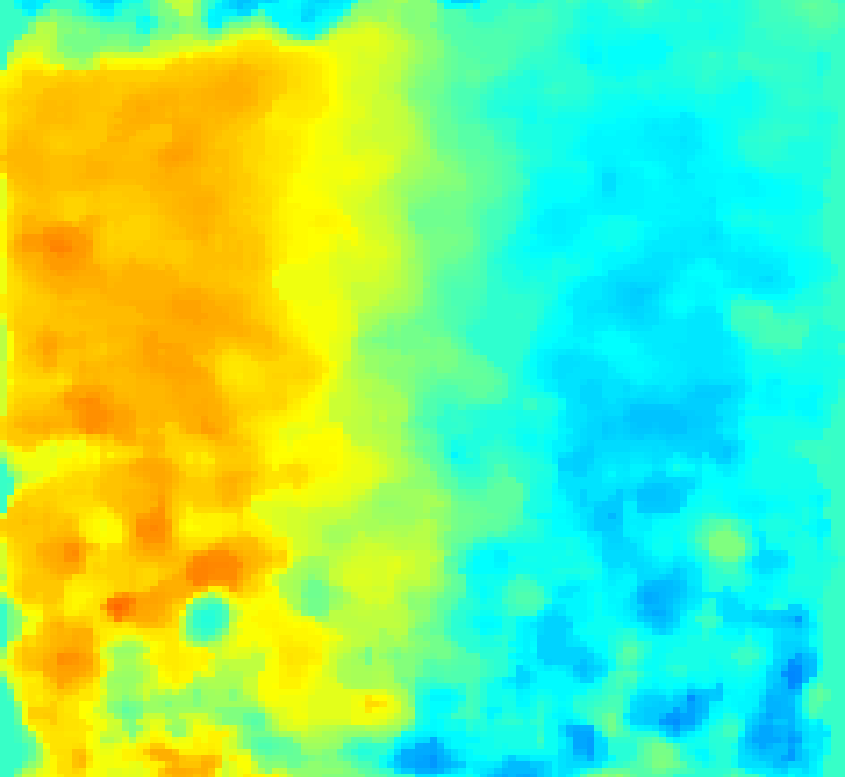}};
        \draw[draw=black] (0.25*3,0.5+0.8*3) rectangle ++(3,1.5);
        \node[anchor=south west,inner sep=0] at (0.25*2,0.5+0.8*2) {\includegraphics[height=15mm,width=30mm]{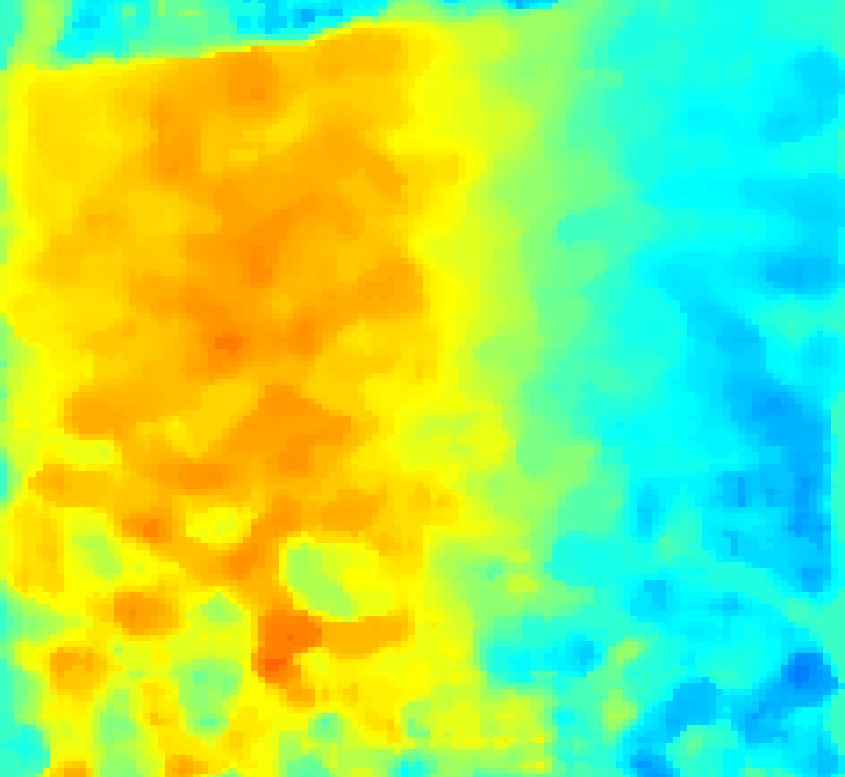}};
        \draw[draw=black] (0.25*2,0.5+0.8*2) rectangle ++(3,1.5);
        \node[anchor=south west,inner sep=0] at (0.25*1,0.5+0.8*1) {\includegraphics[height=15mm,width=30mm]{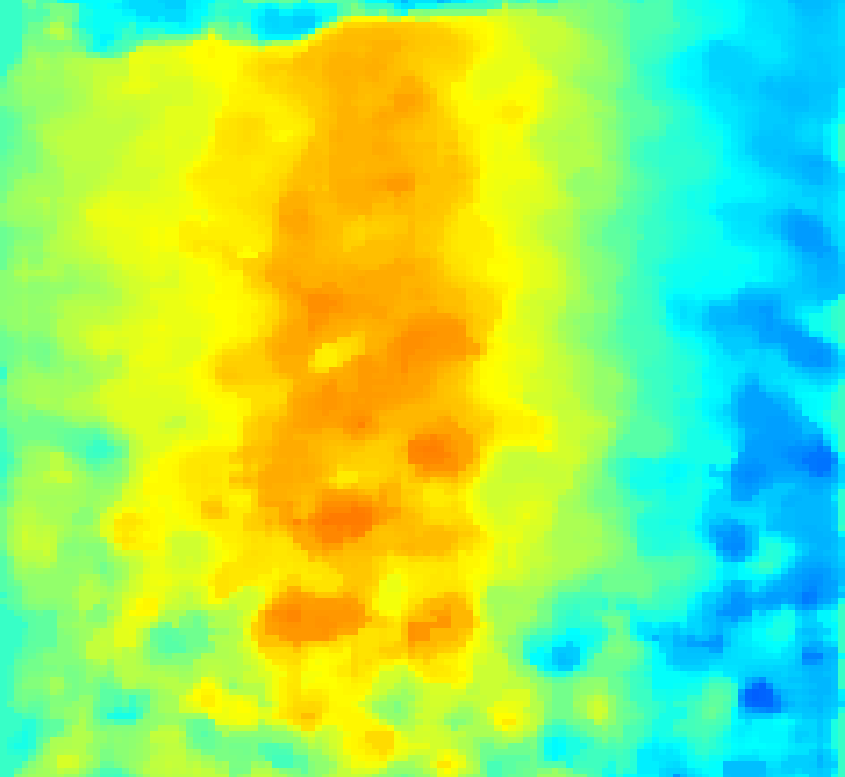}};
        \draw[draw=black] (0.25*1,0.5+0.8*1) rectangle ++(3,1.5);
        \node[anchor=south west,inner sep=0] at (0.25*0,0.5+0.8*0) {\includegraphics[height=15mm,width=30mm]{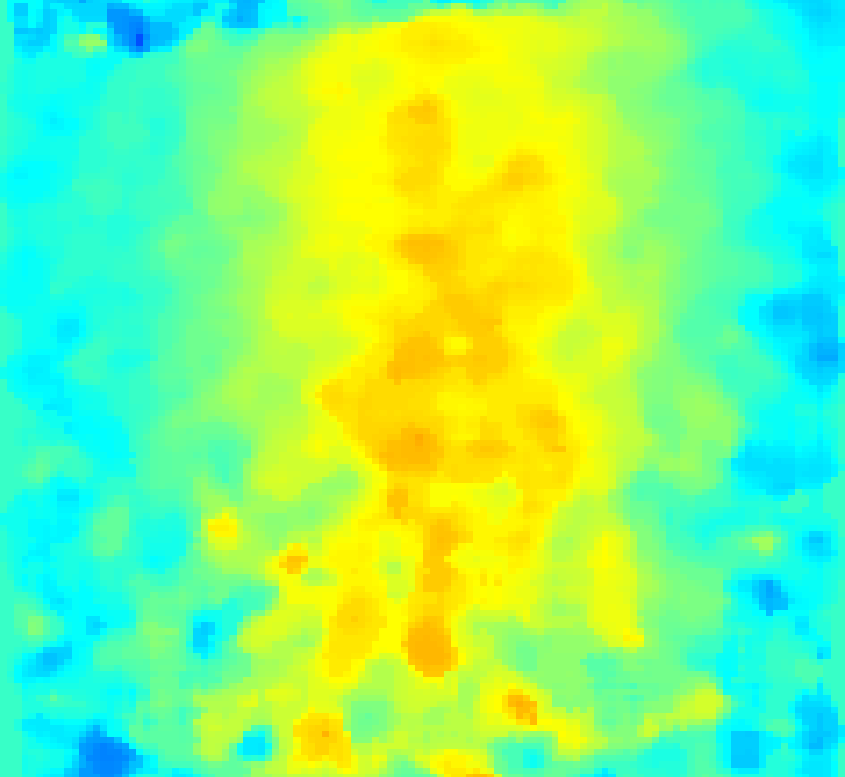}};
        \draw[draw=black] (0.25*0,0.5+0.8*0) rectangle ++(3,1.5);
        
        \draw[-stealth, thick, draw=black] (3.25,0.1+0.5) -- (3.25+0.25*3,0.1+0.5+0.8*3);
        \node at (3.7,1.5) [align=center, anchor=west, rotate=72] {\scriptsize Time};
        \node[anchor=center] at (1.5,0.3){\scriptsize $\mathbb{R}^{w \times h \times t}$};
        \node[anchor=south west,inner sep=0] at (6-0.25,0) {\includegraphics[height=15mm,width=30mm]{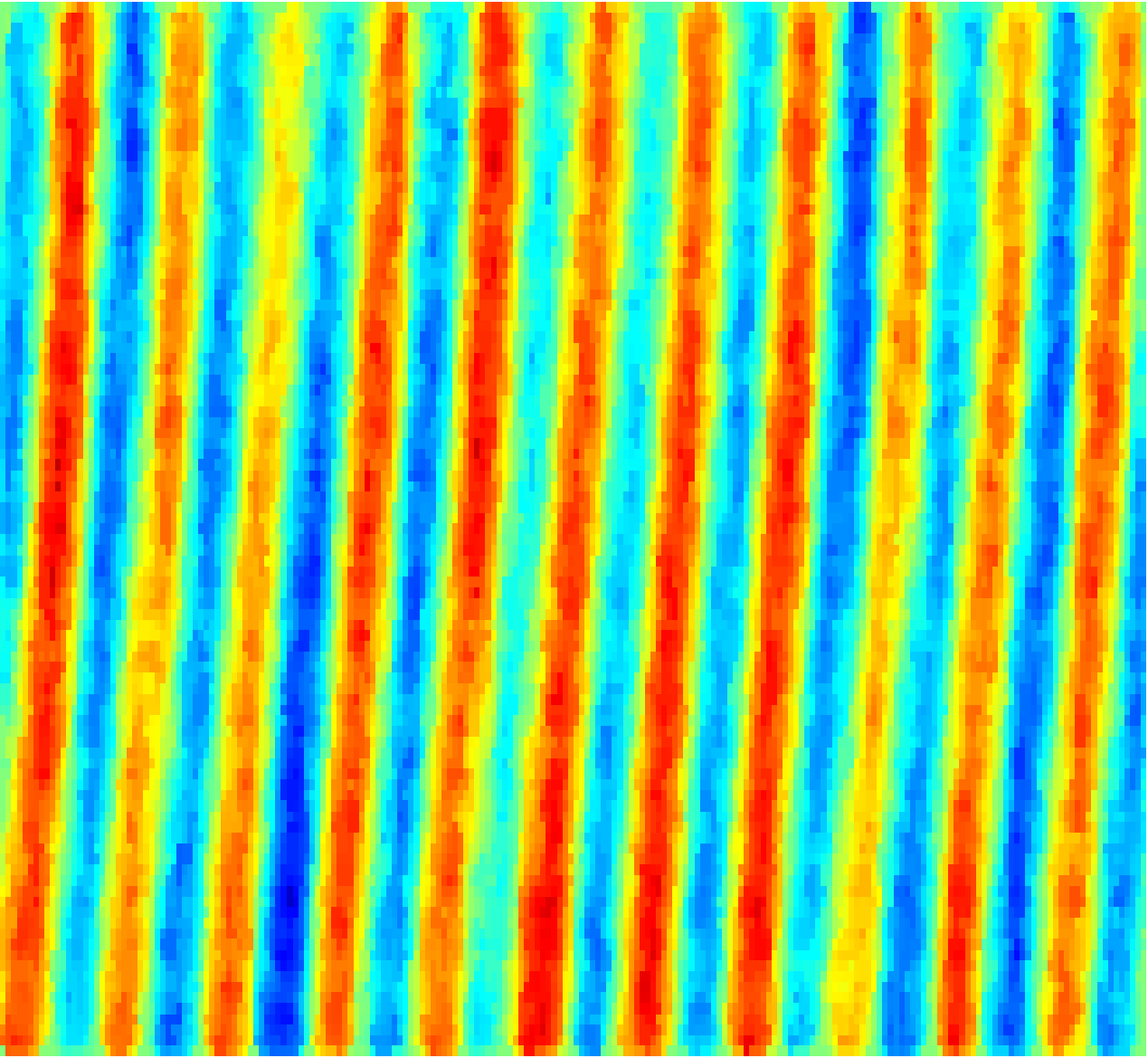}};
        \node[anchor=south west,inner sep=0] at (10,0 ){\includegraphics[height=15mm,width=30mm]{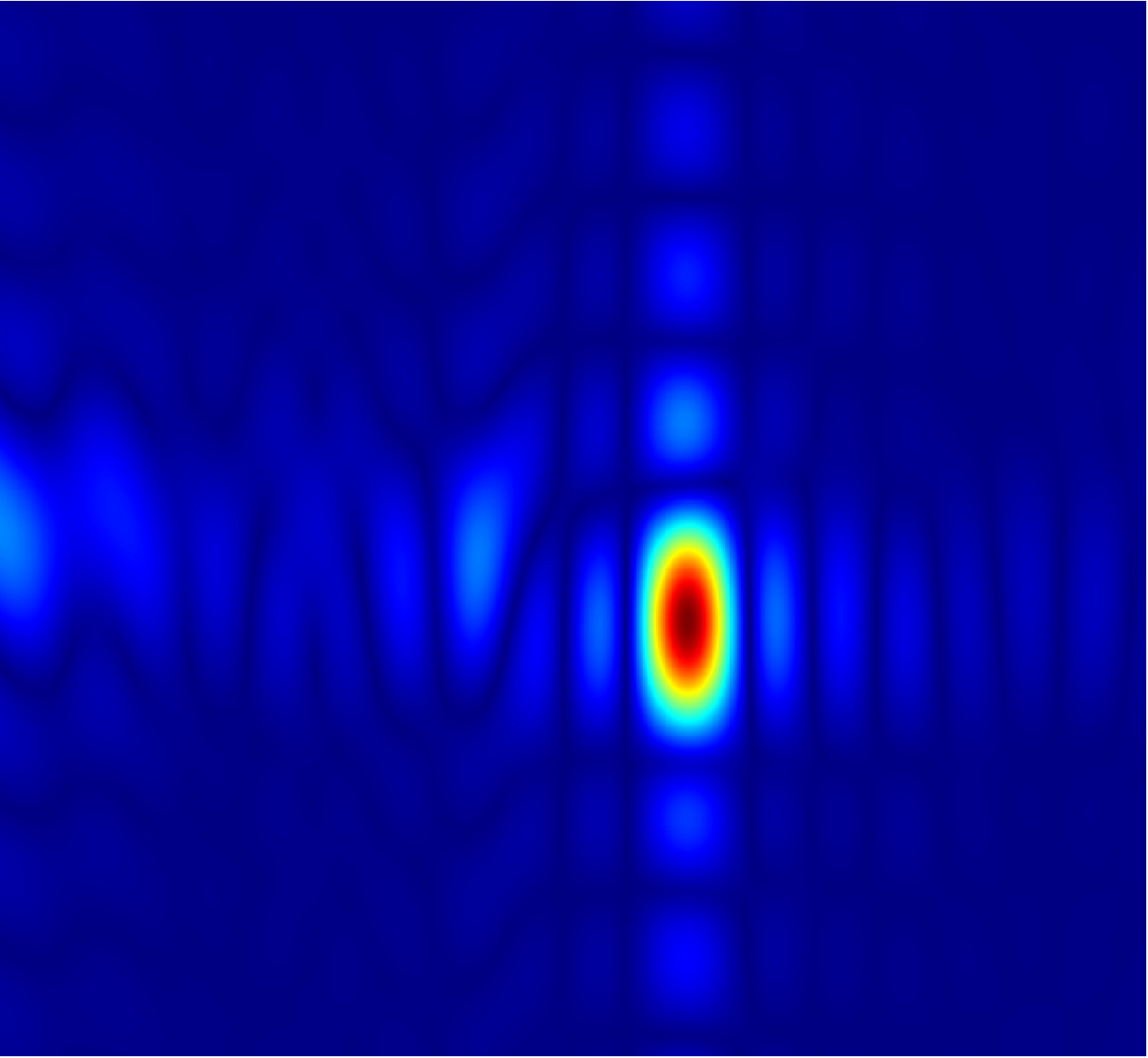}};

        \node at (13,1) [align=center, anchor=west]{\scriptsize $v_{FFT}$};
        
        \node at (14,0.75) [align=center, anchor=west]{\scriptsize Material \\[-1mm]\scriptsize Model};
        \node at (15.8,0.75) [align=center, anchor=west]{\scriptsize Elasticity \\[-1mm]\scriptsize [kPa]};
        \draw[-stealth, thick, draw=black] (4.5,0.75) -- (5,0.75);
        \draw[-stealth, thick, draw=black] (9.5-0.12-0.2,0.75) -- (9.5-0.12+0.2,0.75);
        \node[anchor=center] at (9.5-0.12,0.75+0.3){\scriptsize FFT};

        \draw[-stealth, thick, draw=black] (13.5-0.2,0.75) -- (13.5+0.2,0.75);
        \draw[-stealth, thick, draw=black] (15.5-0.2,0.75) -- (15.5+0.2,0.75);
        \node at (7.5-0.25,1.4) [align=center, anchor=south]{\scriptsize Space-Time Map};
        \node at (11.5,1.4) [align=center, anchor=south]{\scriptsize Frequency Domain};
        \node[anchor=center] at (4.75,0.75+0.3){\scriptsize $\mathbb{R}^{w \times t}$};

        \node[anchor=south west,inner sep=0] at (7-1.7,1.9)  {\includegraphics[height=25mm, width = 114mm]{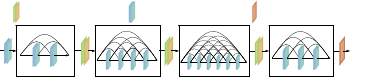}};
        \node at (5.8+0.1,4.0) [align=left, anchor=west]{\scriptsize Transition Layers\\[-1mm] \scriptsize Conv 1x1x1 + Pool Stride 2};
        \node at (9.2+0.2,4.0) [align=left, anchor=west]{\scriptsize Dense Layers\\[-1mm] \scriptsize Conv 3x3x3 + Stride 1};
        \node at (12.8+0.2,4.0) [align=left, anchor=west]{\scriptsize Fully Connected Layer};
        \node at (15.8,2.75) [align=center, anchor=west]{\scriptsize Elasticity \\[-1mm]\scriptsize [kPa]};
        
        \node[anchor=center] at (4.75,3.2){\scriptsize $\mathbb{R}^{w_{l} \times h_{l} \times t_{l}}$};
        \draw[-stealth, thick, draw=black] (4.5,2.75) -- (5,2.75);
        
        \draw[-, thick, draw=white] (0,4.4) -- (5,4.4);
                
        
        \end{tikzpicture}

       \caption{
		\textbf{Data Processing:} The OCT phase signal is cropped below the surface and each 3D spatio-temporal shear wave dataset $\mathbb{R}^{w \times h \times t}$ is processed using conventional phase velocity estimation and our learning based approach. Using spatio-temporal deep learning (upper right), sample elasticity is directly estimated in an end-to-end fashion from a temporal window of $t_{l}=16$ frames. We also downsample the spatial dimension to $w_{l}=h_{l}=64$ to reduce computational effort. For conventional elasticity estimates (lower right), we create space-time maps for each dataset ($w=128$, $t=208$) and estimate shear wave velocity using Fourier domain analysis. We map wave velocity to the Young's modulus using a linear material model.}	
    \label{fig:dataProcessing}
\end{figure*}

\begin{figure*}[tb!]
   \subfloat[Predicted Elasticity]{
       \includegraphics[height=60mm]{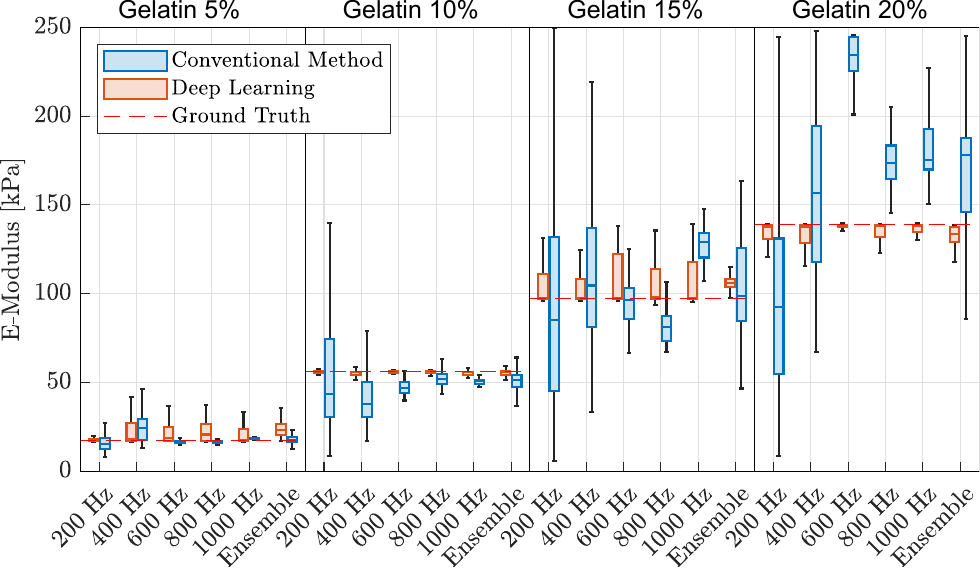}
   \label{fig:bxplt_gel}
    }
   \hfill
      \subfloat[Effect of Damping]{
      \includegraphics[height=60mm]{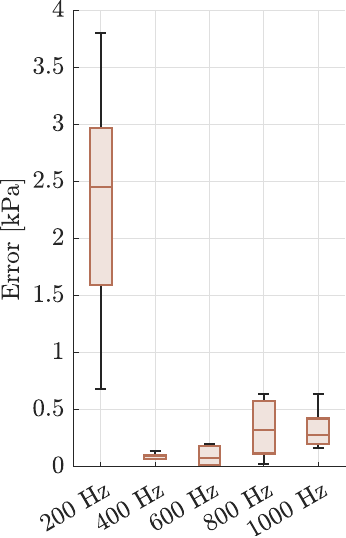}
   \label{fig:bxplt_damp}
    }\hfill
   \subfloat[Frequency]{
      \includegraphics[height=60mm]{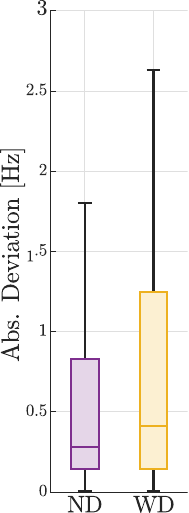}
   \label{fig:bxplt_freq}
    }
   \caption{\textbf{Experimental Results:} (a) Elasticity estimates with deep learning (red) and the conventional method (blue). The dashed red line indicated the target elasticity evaluated by uniform compression tests. (b)  \textcolor{black}{The error in elasticity predictions} due to the damping along the instrument shaft in the laparoscopic trainer. (c) Difference between excitation frequency and measured frequency with damping (WD) and with no damping (ND).}
   \label{fig:bxplt}
\end{figure*}

\subsection{Spatio-Temporal Deep Learning}
The acquired 3D shear wave data consists of two spatial and one temporal dimension for the processing. Spatio-temporal Convolutional Neural Networks (CNNs)~\cite{tran2015learning} with 3D convolutions have shown excellent results~\cite{neidhardt20214d, neidhardt2020deep} by jointly learning spatial and temporal information. For efficient data processing, we resize each frame resulting in 64 \texttimes\ 64~pixels along the lateral and depth axis, respectively. Our model architecture (Fig.\ref{fig:dataProcessing}) is based on a densely connected CNN (DenseNet)~\cite{Huang.2017}. After an initial convolutional layer, we employ four DenseNet blocks and with two, four, six, and three layers per block, respectively. We use average pooling in the transition blocks and a growth factor of 6. A final fully connected layer directly outputs elasticity estimates given the 3D input sequence. We train independent networks for each excitation frequency using a mean squared error (MSE) loss. We split our data based on the different phantoms and select one independent phantom from each gelatin concentration as our test set. To make our results more robust, we conduct a nested cross-validation scheme such that every phantom is part of the test set once. For each of these five outer folds, we perform a four-fold cross-validation across the remaining samples. This results in a data split of 70:20:10 for training, testing, and validation respectively. Thereby, we train 20 networks for each frequency leading to 100 networks trained in total. During training, we randomly crop sequences of 16 successive frames along the temporal dimension. During testing, we apply a moving window along the temporal dimension and report the mean of all predictions from one dataset. We apply early stopping to reduce training times. While we train solely on gelatin but additionally evaluate our approach on soft tissue data.

\subsection{Conventional Elasticity Estimation}
For comparison to our deep-learning approach, we additionally consider conventional phase velocity estimation as depicted in Fig.\ref{fig:dataProcessing}. We estimate the velocity of the shear waves in the frequency domain similar to Beuve et al.~\cite{Beuve.2021b}. First, we derive Space-Time maps from a given shear wave dataset by estimating the mean along the depth axis $h$, resulting in an image of size of 118 \texttimes\ 208~pixels along the lateral $w$ and temporal $t$ axis. Second, we apply a 2D discrete FFT along the image axis to estimate the $k$-space representation. To further reduce background noise we remove pixels with $<10\%$ of the overall maximum amplitude in the k-space. Next, we estimate the maximum amplitude in the k-space for each temporal frequency $f$ and compute the phase velocity $v_{FFT} = {f}/{k}$ with the wavenumber $k$. Estimates not within $\SI[]{1}{\meter\per \second}<v_{FFT}>\SI[]{10}{\meter\per \second}$ are considered failed estimates, which represent the typical elasticity range of soft tissue~\cite{fink2010multiwave}. Finally, we apply a linear material model to estimate the E-Modulus from the shear wave velocity with the relation~\cite{sarvazyan2013acoustic} $ E_{ToF} = q \cdot \rho \cdot 2(1+\nu) \cdot v_{FFT}^{2}$ with a Possion's ratio $\nu=0.5$ and a density $\rho=\SI{1000}{\kilo\gram\per\cubic\metre}$. To account for constant errors between our estimates and our ground truth Young's modulus labels we estimate a linear scaling factor $q=0.84$ by minimizing the offset.\\

\begin{table*}[htb]
    \caption{Mean absolute error of predictions in \SI[]{}{\kilo \pascal} on all gelatin phantoms with different excitation frequencies.}
    \label{tab:MAE_Table_1}
    \centering
    \begin{tabular}{l c D{,}{\pm}{4.4} D{,}{\pm}{4.4} D{,}{\pm}{4.4} D{,}{\pm}{4.4} D{,}{\pm}{4.4} D{,}{\pm}{4.4} }
        \toprule
        Method
        & Error Metric
        & \multicolumn{1}{c}{\SI{200}{Hz}}
        & \multicolumn{1}{c}{\SI{400}{Hz}}
        & \multicolumn{1}{c}{\SI{600}{Hz}}
        & \multicolumn{1}{c}{\SI{800}{Hz}}
        & \multicolumn{1}{c}{\SI{1000}{Hz}}
        & \multicolumn{1}{c}{Ensemble}\\
        \midrule
        Conventional 
        &MAE
        &34.06,34.81
        &26.53,26.18
        &25.16,37.57
        &16.00,15.85
        &22.66,19.85
        &19.27,23.50 \\
        &RMSE
        &\multicolumn{1}{c}{48.67}
        &\multicolumn{1}{c}{37.26}
        &\multicolumn{1}{c}{45.18}
        &\multicolumn{1}{c}{22.51}
        &\multicolumn{1}{c}{30.11}
        &\multicolumn{1}{c}{30.37}\\
        Deep Learning
        &MAE
        &6.42,13.45
        &6.73,11.97
        &6.25,12.61
        &7.58,13.63
        &6.66,12.63
        &6.29,4.99\\
        &RMSE
        &\multicolumn{1}{c}{14.89}
        &\multicolumn{1}{c}{13.72}
        &\multicolumn{1}{c}{14.06}
        &\multicolumn{1}{c}{15.58}
        &\multicolumn{1}{c}{14.27}
        &\multicolumn{1}{c}{8.02}\\
        \bottomrule
    \end{tabular}
\end{table*}

\section{RESULTS}
\begin{figure}[tb]
    \includegraphics[width=0.9\linewidth]{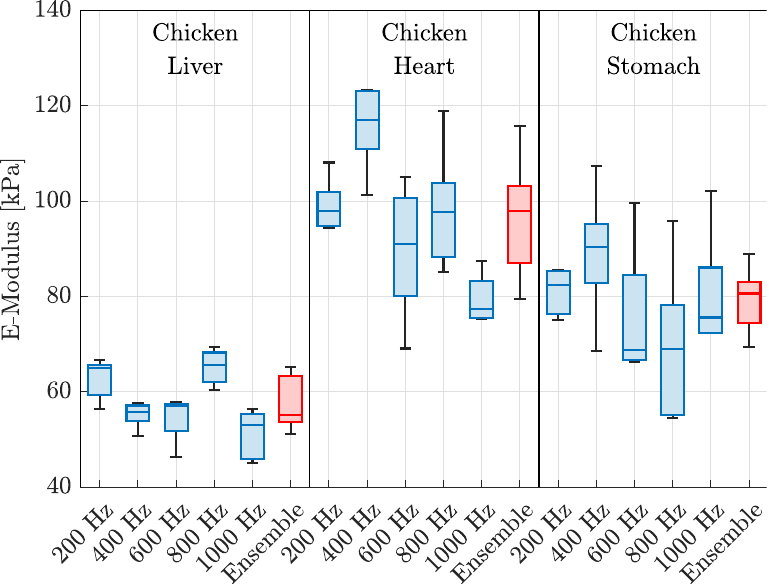}

    \caption{
		\textbf{Ex-vivo Soft Tissues:} Elasticity estimates for chicken liver, stomach and heart tissue. We additionally consider the ensemble by taking the mean over the models trained for each of the five frequencies.
  }
    \label{fig:chicken}
\end{figure}

Example data of a shear wave excited using our modified da~Vinci instrument can be seen in Fig.\ref{fig:dataProcessing}. Visual inspection of the acquired datasets confirmed wave propagation for all datasets and samples, with varying imaging noise dependent on material and excitation frequency. All 3D shear wave datasets for each sample were processed using the conventional approach and our spatio-temporal deep learning based elasticity estimation. The mean absolute error (MAE) and the root mean squared error (RMSE) averaged for each excitation frequency can be seen in Table~\ref{tab:MAE_Table_1} for both methods. Conventional processing displays significant variation in elasticity estimation and is outperformed by the deep learning-based approach, which enables a more accurate and more robust estimation of the elasticity independent of the excitation frequency. The elasticity estimates for each excitation frequency and sample stiffness are shown in Fig.\ref{fig:bxplt_gel} for both approaches. For the conventional approach, it can be seen that the variance increases with elasticity and decreases with frequency, resulting in an MAE of \SI{19.27(2350)}{\kilo\Pa} for the ensemble from all frequencies. In contrast, model predictions for the learning-based approach show good agreement with the acquired ground truth independent of excitation frequency and we report a reduced MAE of \SI{6.29(499)}{\kilo\Pa}. %
We further investigate the effect of damping on the wave excitation with the instrument in the laparoscopic trainer. We report the error when damping is added (Fig.\ref{fig:bxplt_damp}). We observe only marginal differences in the elasticity estimates for excitation with and without damping, with a maximum mean deviation of \SI{2.31}{\kilo\Pa} for the lowest excitation frequency. We further compare the excitation frequency with the measured dominant frequency component in the frequency domain (Fig.\ref{fig:bxplt_freq}). The median deviation between excitation and measured shear wave frequency is \SI{0.28(153)}{\Hz} and \SI{0.41(164)}{\Hz} with and without damping, respectively. In Fig.\ref{fig:chicken}, the elasticity estimates of ex-vivo soft tissue samples are depicted. In average we report an elasticity of \SI{57.73(595)}{\kilo\Pa}, \SI{96.39(1337)}{\kilo\Pa} and \SI{79.27(718)}{\kilo\Pa} for chicken liver, heart and stomach, respectively.

\section{DISCUSSION}

Our experiments demonstrate the successful excitation of shear waves in phantom and ex-vivo tissue samples directly via the modified da~Vinci instrument. Further, we evaluate the effect of damping on the instrument shaft, as presented in a clinical minimal invasive intervention. We observe mean deviations of less than \SI{1}{\Hz} between excitation frequency and measured frequency inside the tissue, both with and without additional damping in the laparoscopic trainer. Our modified instrument consequently enables a safe and reliable approach for wave excitation from outside the patient. In contrast, previously proposed probes for wave excitation involve piezoelectric elements at the tip of dedicated tools~\cite{Neidhardt.2023, Karpiouk.2018} with  \textcolor{black}{high voltage cables running inside the patient}. Air pulse excitation is a promising non-contact alternative and allows highly focused wave excitations \cite{Zvietcovich.2022}. However, current designs have limited bandwidth and are impractical due to their location inside the patient, close to the soft tissue.

Our experiments confirm that our approach enables the differentiation of materials based solely on the propagating shear wave field (Fig.\ref{fig:bxplt}). The proposed spatio-temporal deep learning enables accurate elasticity estimates and outperforms conventional processing with an MAE of \SI{6.29(499)}{\kilo\Pa} compared to \SI{19.27(2350)}{\kilo\Pa}, respectively. The conventional approach results in large deviations for estimates with different excitation frequencies from the same elasticity, especially for lower excitation frequencies and increased sample elasticity. In contrast, the trained models are more robust with respect to the excitation frequency and sample stiffness, especially when considering ensemble predictions. We further investigate if damping of the instrument's shaft, as given in a minimally invasive intervention, effects elasticity prediction. Considering Fig.\ref{fig:bxplt_damp}, damping effects are marginal on elasticity estimates with a maximum mean estimation offset of \SI{2.71(80)}{\kilo\Pa} for excitation frequencies of \SI{200}{\Hz}. Consequently, the excitation frequency can be modulated to provide more accurate elasticity estimates.

We additionally evaluate our trained models on ex-vivo soft tissue and demonstrate that different tissue types can be differentiated based on their elasticity, with particularly good contrast between softer liver and harder heart and stomach muscle tissue. It stands out, that while training the models was solely performed on gelatin phantoms the models are able to generalize to soft tissue. The results therefore show successful shear wave excitation in soft tissue and indicate that the wave field has similar features as in gelatin phantoms. \textcolor{black}{ Overall, chicken heart and stomach are in the range of muscle tissue reported in the literature \SI{98(10)}{\kilo\Pa} \cite{roux2023model}. However, a comparison to quantitative values in the literature is limited due to the not applicable ground truth elasticity of our samples as well as the varying age and the general inhomogeneity of soft-tissue samples, as well as different presumptions for elasticity estimates, e.g., the material model. In this context, our model can also be applied for respective downstream tasks, e.g., surgical navigation, tissue classification, or tactile feedback. }

A refined instrument design will be required for clinical application, especially for medical autoclaving. Consequently, 3D printed parts ideal for fast prototyping would need to be replaced by metal alloys and electric wiring could be directly integrated into the existing da~Vinci electrical adapter. \textcolor{black}{In practice, the imaging of the tissue could be performed similarly to existing OCT handheld probes \cite{benboujja2016intraoperative}. Volumetric imaging could be used to detect wave propagation along multiple spatial dimensions~\cite{neidhardt20214d}. We note that the imaging depth of OCE is limited, but its high resolution makes it particularly useful for highly local elasticity estimation.}

Currently, our study also only considers excitation with a rigid instrument tip position. Future studies will consider wave excitation with simultaneous actuation of the instrument tip, e.g., push and grabbing tasks.

In conclusion, we have demonstrated shear wave excitation directly via a modified da~Vinci instrument and shown subsequent differentiation of different phantoms and soft tissue types based on their elastic properties and considering friction at the instrument shaft. The availability of quantitative elasticity estimates during minimal invasive RAS offers multiple possibilities including the assistance of the surgeon during navigation, classifying tissue types or enabling quantitative haptic feedback.

\bibliographystyle{IEEEtran}
\bibliography{refs}

\end{document}